\documentclass[
    onecolumn,
    preprintnumbers,
    prl,
    showpacs
]{revtex4}
\usepackage{graphicx}%
\usepackage{dcolumn}
\usepackage{amsmath}
\usepackage[normalem,normalbf]{ulem}
\usepackage{amssymb}
\usepackage{bm}
\usepackage{enumerate}

\begin{document}
\title{New variational perturbation theory based on $q-$deformed
    oscillator}
\author{Hyeong-Chan Kim}
\email{hckim@phya.yonsei.ac.kr}
\author{Jae Hyung Yee}%
\email{jhyee@phya.yonsei.ac.kr}
\affiliation{Department of Physics, Yonsei University, Seoul
120-749, Korea.
}%
\author{Sang Pyo Kim}
\email{sangkim@kunsan.ac.kr}
\affiliation{Department of Physics, Kunsan National University,
Kunsan 573-701, Korea.
}%
\date{\today}%
\bigskip
\begin{abstract}
\bigskip

A new variational perturbation theory is developed based on the
$q-$deformed oscillator. It is shown that the new variational
perturbation method provides 200 and 10 times better accuracy for
the ground state energy of anharmonic oscillator than the Gaussian
and the post Gaussian approximation, respectively, for weak
coupling.

\end{abstract}
\maketitle

There have been many proposals to establish methods of extracting
nonperturbative information from quantum mechanical and quantum
field theoretic systems, such as $1/N$ expansion and methods based
on Dyson-Schwinger equation~\cite{justin}. One of the successful
ways to obtain nonperturbative information is the variational
perturbation theory~\cite{stevenson,jackiw}. Many variants of the
variational perturbation theory based on the Gaussian approximation
have been used to study various aspects of physical
systems~\cite{kleinert,lee,bak,okopinska,okopinska1,rakhimov,amore}.
Although the variational perturbation theory provides one with
systematic correction terms to the variationally determined
approximation, it has a limitation that only the Gaussian wave
function(al) may be used as a variational trial wave function(al)
for the most physical systems. In this paper we attempt to establish
a new variational perturbation theory based on the $q-$deformed
oscillator~\cite{bonatsos,macfarlane}, which provides a better
approximation than that based on the Gaussian approximation.

Quantum anharmonic oscillator has been frequently used in developing
various approximation methods in quantum mechanics and quantum field
theory~\cite{diney,justin,kleinert,lee,bak}. To develop a new
variational perturbation method we will consider the anharmonic
oscillator given by the Hamiltonian,
\begin{eqnarray} \label{H:anhar}
\hat{H}= \frac{\hat p^2}{2}+ \frac{\omega^2}{2}\hat x^2+
\frac{\lambda}{4} \hat{x}^4.
\end{eqnarray}

We first illustrate some essential procedures of the variational
perturbation theory based on the Gaussian approximation. We start by
writing the Hamiltonian~(\ref{H:anhar}) as the sum of the Gaussian
Hamiltonian, $H_G$, and the perturbation term, $V_I$, as
\begin{eqnarray} \label{H:VI}
\hat{H}= \hat{H}_G+ V_I,
\end{eqnarray}
where
\begin{eqnarray} \label{H:G}
\hat H_G \equiv \frac{1}{2}\hat{p}^2+ \frac{\Omega_G^2}{2}
\hat{x}^2,~~V_I = \frac{\omega^2-\Omega_G^2}{2}
\hat{x}^2+\frac{\lambda}{4} \hat{x}^4 .
\end{eqnarray}
Since the Gaussian Hamiltonian is that of a simple harmonic
oscillator, it can be written in terms of the creation and the
annihilation operator, $\displaystyle \hat
A_G=\sqrt{\frac{\Omega_G}{2\hbar}}\hat x+ \frac{i}{\sqrt{2 \hbar
\Omega_G}} \hat p$, in the quadratic form:
\begin{eqnarray} \label{H_G:A}
\hat H_G =\frac{\hbar \Omega_G}{2}\left(\hat{A}_G^\dagger
\hat{A}_G+\hat{A}_G\hat{A}_G^\dagger\right).
\end{eqnarray}
Moreover, the Lie algebra, $[\hat{A}_G,\hat{A}_G^\dagger]=1$, allows
one to define a Fock space consisted of the ground state
$|0\rangle_G$ defined by,
\begin{eqnarray} \label{0:G}
\hat{A}_G|0\rangle_G=0,
\end{eqnarray}
and the excited states, $|n\rangle_G$, generated by successively
acting $A^\dagger_G$ on $|0\rangle_G$.

The next step is the variational procedure with respect to
$\Omega_G$ so that the energy of the Hamiltonian~(\ref{H:VI}) with
respect to the Gaussian ground state is minimized. Some of the
nonperturbative effects of the Hamiltonian~(\ref{H:anhar}) are
amalgamated to $\hat H_G$ by this process, which makes the
correction term, $V_I$, to be an order of $\xi_G$ (defined below)
smaller than the Gaussian Hamiltonian. The $V_I$ term may be written
as in Ref.~\cite{bak} as
\begin{eqnarray} \label{VI}
V_I= \frac{\xi_G\hbar\Omega_G}{2}\left[-1 +\frac{1}{3} \sum_{k=0}^4 ~\left(\begin{tabular}{c} 4\\
$k$\end{tabular}\right) \hat{A}_G^{\dagger k} \hat{A}_G^{4-k}
\right] ,
\end{eqnarray}
where $\displaystyle \xi_G\left(=\frac{3\lambda \hbar}{2
\Omega_G^3}\right) <1$ for any value of $\lambda$ due to the
variational gap equation, $\displaystyle
1-\xi_G=\frac{\omega^2}{\Omega_G^2}$. In this sense, the Gaussian
Hamiltonian $\hat H_G$ describes a truncated form of $\hat H$ up to
$O(\xi_G^{~0})$.

The final step in the variational perturbation theory is to apply
the conventional perturbation method to the Hamiltonian~(\ref{H:VI})
by using the complete set of states, $\{|n\rangle_G\}$. In an
operator method, the so called Liouville-von Neumann approach, this
is achieved by defining the $1^{\rm st}$ order annihilation
operator, $\hat A_{(1)}= \hat A_G+ \xi_G \hat B$, so that the
Hamiltonian $\hat H$ can be factorized as $\displaystyle \hat H
=\frac{\hbar \Omega}{2}\left(\hat{A}_{(1)}^\dagger
\hat{A}_{(1)}+\hat{A}_{(1)}\hat{A}_{(1)}^\dagger\right)$, up to the
first order in $\xi_G$~\cite{bak} .  It was shown that these
operators satisfy the (real) $q-$deformed algebra~\cite{bonatsos}
with deformation parameter, $q^2=1+\xi_G/2$. This algebra leads to
another Fock space determined by the $1^{\rm st}$ order creation
operator, $\hat{A}^\dagger_{(1)}$, which defines the $1^{\rm st}$
order states $|n\rangle_{(1)}$.

The fact that the first order perturbative correction to the
Gaussian approximation is described by the $q-$deformed oscillator
indicates that one may establish a better variational perturbation
theory by using the $q-$deformed oscillator as the basis for the
variational method. To develop the variational perturbation method
based on the $q-$deformed oscillator, we start by separating the
Hamiltonian~(\ref{H:anhar}) into the $q-$deformed part, $\hat H_q$,
and the perturbation term, $V_I'$:
\begin{eqnarray} \label{H:Hq:VI}
\hat H= \hat H_q + V_I',
\end{eqnarray}
where $V_I'= \hat{H}-\hat{H}_q$, and
\begin{eqnarray} \label{H:q}
\hat{H}_q \equiv
\frac{\hbar\bar{\Omega}}{2}\left(\hat{a}_q\hat{a}_q^\dagger +
\hat{a}_q^\dagger \hat{a}_q\right)
\end{eqnarray}
is the Hamiltonian of the $q-$deformed oscillator~\cite{bonatsos},
with $\hat a_q$ and $\hat{a}_q^\dagger$ satisfying the algebra,
\begin{eqnarray}\label{q:deformed}
[\hat{a}_q, \hat{a}_q^\dagger]=1+ \epsilon
  \hat{a}_q^\dagger \hat{a}_q ,
\end{eqnarray}
where we use this specific form of $q-$algebra, since this is
satisfied with the perturbative anharmonic oscillator~\cite{bak}.
The $q-$ground state is defined by
\begin{eqnarray}\label{0:q}
\hat{a}_q |0\rangle_q=0,
\end{eqnarray}
and the $q-$excited states can be generated by successively acting
$\hat{a}_q^\dagger$ to the ground state~(\ref{0:q}).

For calculational simplicity, we introduce dimensionless operator
$\hat H_m^n$ of order $O(\hbar^0)$:
\begin{eqnarray} \label{Hmn}
\hat H_m^n= \left(\frac{\hbar}{2 \Omega_q}\right)^{-\frac{m}{2}}
    \left(\frac{\Omega_q\hbar}{2}\right)^{-\frac{n}{2}}\frac{\hat
    x^m \hat p^n+ \hat p^n \hat x^m }{2}.
\end{eqnarray}
Similar operators are defined in Ref.~\cite{bender}. The explicit
expression for $\hat a_q$, which satisfies the
algebra~(\ref{q:deformed}) and make the Hamiltonian~(\ref{H:q}) to
be (\ref{Hq3}) is obtained to $O(\epsilon^2)$ in Appendix A,
\begin{eqnarray} \label{a:results}
\hat a_q &=&u^*\left\{i\hat{H}_0^1+
    \left(1+\frac{\epsilon^2}{4}\right)\hat{H}_1^0
    +\frac{\epsilon(1-\epsilon/2)
    }{4}\left[\hat{H}_1^2-i\hat{H}_2^1+\frac{2}{3} \hat{H}_3^0\right]
    +\frac{\epsilon^2}{32}\left[-i \hat{H}_2^3- \hat{H}_3^2-i\frac{4}{3}
    \hat{H}_4^1+\frac{8}{5} \hat{H}_5^0 \right] \right\}+\cdots,
\end{eqnarray}
where $\displaystyle u^* = \frac
    {1}{2}\left(1-\frac{1}{4}\epsilon-\frac{9}{32}\epsilon^2
 \right)$.  After inverting the expression for
$\hat a_q$ to get $\hat{H}_1^0$ as a function of $\hat{a}_q$ and
$\hat{a}_q^\dagger$, and taking the expectation value with respect
to the $q-$ground state~(\ref{0:q}), we obtain
\begin{eqnarray} \label{exp}
 \langle \hat{H}_2^0\rangle_q=
1-\epsilon+\frac{13}{24}\epsilon^2, ~~\langle
\hat{H}_4^0\rangle_q=3-7\epsilon+\frac{33}{4}\epsilon^2,
~~~\langle\hat{H}_6^0\rangle_q=15 .
\end{eqnarray}

Note that we have introduced two variational parameters $\bar\Omega$
and $\epsilon$.  To establish the variational perturbation theory
based on the $q-$Hamiltonian, we need to express the
$q-$Hamiltonian~(\ref{H:q}) in terms of the phase space variables,
$\hat x$ and $\hat p$. By demanding that the $q-$Hamiltonian is
frictionless (no $\hat{p} \hat{x}+\hat{x}\hat{p}$ term), the
$q-$Hamiltonian may be written as, up to $O(\epsilon^2)$,
\begin{eqnarray} \label{q-ham}
\hat{H}_q &=&\frac{\hbar \Omega_q}{4}\left[\hat{H}_2^0+\hat{H}_0^2+
   \frac{\epsilon}{3}\left(1-\frac{\epsilon}{2}\right)\hat{H}_4^0
   -\frac{g_3\epsilon^2}{4}
     \hat{H}_6^0 +\cdots\right] .
\end{eqnarray}
The parameters in~(\ref{H:q}) and (\ref{q-ham}) are determined in
the appendix A and $\displaystyle g_3=-\frac{23}{45}$.

Note that, unlike the case of the variational perturbation theory
based on the Gaussian approximation, the $q-$Hamiltonian $\hat H_q$
is not expressed in a closed form in the phase space, but is written
as a series in $\epsilon$. Since we are to compute up to the $1^{\rm
st}$ order correction to the variational result, it is enough to
write $\hat{H}_q$ up to $O(\epsilon^2)$ as in Eq.~(\ref{q-ham}).

The Hamiltonian $\hat H$ of Eq.~(\ref{H:anhar}) is then written as,
$\hat{H}= \hat{H}_q+ V_I'$, up to $O(\epsilon^2)$, where
\begin{eqnarray}\label{VI'}
V_I'=\frac{\hbar\Omega_q}{4}\left[\left(\frac{\omega^2}{\Omega_q^2}
         -1\right)
      \hat{H}_2^0 +\frac{\xi-2\epsilon+\epsilon^2}{6}\hat{H}_4^0
         +\frac{g_3 \epsilon^2}{4} \hat{H}_6^0 +\cdots\right],
\end{eqnarray}
with $\displaystyle \xi=\frac{3\lambda \hbar}{2\Omega_q^3}$. Taking
expectation value of (\ref{H:Hq:VI}) with respect to the $q-$ground
state~(\ref{0:q}), the energy expectation value of the anharmonic
oscillator becomes
\begin{eqnarray} \label{E:0}
_q\langle 0 |\hat H | 0 \rangle_q &=&
  \frac{\hbar\Omega_q}{4}\left[1+\epsilon+\frac{5}{8}\epsilon^2
   +\left(1-\epsilon+\frac{13}{24}\epsilon^2\right)
       \left(\frac{\xi}{\xi_0}\right)^{2/3}
       +\left(1-\frac{7}{3}\epsilon+\frac{11}{4}\epsilon^2\right)
   \frac{\xi}{2}
\right]+ O(\epsilon^3).
\end{eqnarray}
The variational minimization of the energy expectation
value~(\ref{E:0}), with respect to $\Omega_q$ and $\epsilon$, leads
to the gap equations,
\begin{eqnarray} \label{gap}
\left(1-\epsilon+\frac{13}{24}\epsilon^2\right)
   \left(\frac{\xi}{\xi_0}\right)^{2/3}
  &=&1+
\epsilon + \frac{5}{8} \epsilon^2-\left(1-\frac{7}{3}\epsilon +
\frac{11}{4}\epsilon^2
\right) \xi, \\
\frac{5}{4}\left[1+\frac{13}{15}\left(\frac{\xi}{\xi_0}\right)^{2/3}
  +\frac{11}{5}\xi\right]\epsilon&=&\left(\frac{\xi}{\xi_0}\right)^{2/3}
    +\frac{7}{6}\xi-1 , \nonumber
\end{eqnarray}
where $\displaystyle \xi_0=\frac{3\lambda \hbar}{2\omega^3}$. The
gap equations relate $\xi$ to $\epsilon$ as,
\begin{eqnarray} \label{x:ep}
\xi &=& 2\epsilon \left(1- \frac{3\epsilon}{4}+\frac{65
\epsilon^2}{16}\right) f, \\
f&=&\left(
1-3\epsilon-\frac{91\epsilon^2}{8}+\frac{143\epsilon^3}{16}
\right)^{-1}.
\end{eqnarray}
By using Eq.~(\ref{gap}), the ground state energy of the anharmonic
oscillator becomes
\begin{eqnarray} \label{E0:2}
E_0 &=& \frac{\hbar\omega}{2}\left(\frac{\xi_0}{\xi}\right)^{1/3}
  \left[1+
  \epsilon+\frac{5}{8}\epsilon^2 - \left(1-\frac{7}{3}\epsilon
    +\frac{11}{4}\epsilon^2\right)\frac{\xi(\epsilon)}{4}\right]
      +O(\epsilon^3), \nonumber
\end{eqnarray}
where $\xi(\xi_0)$ and $\epsilon(\xi_0)$ are determined by the gap
equations~(\ref{gap}). We present the values of $\epsilon$, $\xi$,
and $E_0$ for several values of the coupling, $\xi_0$, in Table I.
As can be seen in this table, the maximal error of the present
approximation is less than $0.8\%$. Moreover, the accuracy is 200
times and 10 times better than the Gaussian and the post (including
the $2^{\rm nd}$ order perturbative corrections) Gaussian
approximations~\cite{lee}, respectively, for weak coupling. The
maximum value of the dimensionless expansion parameter, $\epsilon
\simeq 0.1717$, is attained as $\xi_0\rightarrow \infty$, which is
distinguishably small compared to the Gaussian expansion parameter,
$\xi_G=1$, at the same limit. This clearly shows that the present
method provides better nonperturbative information than the Gaussian
approximation and $2^{\rm nd}$ order variational perturbation
results based on the Gaussian approximation.

\vspace{.5cm}
\begin{table}
\caption{ground state energy of anharmonic oscillator
\label{ground}}
\begin{tabular}{|c|c|c|c|c|c|c|c|c|}\hline
 &
\multicolumn{2}{c|}{Gaussian} & \multicolumn{2}{c|}{post
Gaussian(2$^{\rm nd}$)}
 & \multicolumn{4}{c|}{present method}    \\
\cline{2-9} $\displaystyle \xi_0$
  & $E_0/(\hbar \omega)$ & error(\%) &$E_0/(\hbar \omega)$ & error(\%) &
    $E_0/(\hbar \omega)$ & error(\%) & $\xi$ &
                        $\epsilon$  \\
\tableline
 0 &    1/2 &0     &  1/2  & 0      & 1/2 &0        &0 & 0 \\
0.06 & 0.507& 0.006& 0.507& -0.0003 &0.507& 0.00003 &0.0599&0.028\\
0.6&   0.560& 0.21 & 0.559&-0.037   & 0.559&0.0083 &0.481& 0.121 \\
6 &    0.813& 1.09 & 0.801&-0.37    & 0.806& 0.307 &1.264& 0.162 \\
60 &   1.531&1.75&   1.49&-0.68     &1.514& 0.628  &1.627& 0.170 \\
600&   3.19&1.95&    3.11&-0.79     &3.15& 0.731   &1.722& 0.171 \\
$\infty$& 0.375 $\xi_0^{1/3}$ & 2.01& 0.365 $\xi_0^{1/3}$& -0.82&
        0.370$\xi_0^{1/3}$&0.766  & 1.75 &0.172 \\
 \hline
\end{tabular}
\end{table}

To complete the $1^{\rm st}$ order perturbative corrections to the
variational result, we need to construct the $1^{\rm st}$ order
creation and annihilation operators which are correct to
$O(\epsilon^2)$. To do this we need to express the
Hamiltonian~(\ref{H:Hq:VI}) as a function of $q-$operators,
$\hat{a}_q$ and $\hat{a}_q^\dagger$. Eq.~(\ref{x:ep}) enables one to
write the coefficients in the potential $V_I'$, Eq.~(\ref{VI'}), as
\begin{eqnarray} \label{summ}
 \frac{\omega^2}{\Omega_q^2}-1 &=&
\frac{\epsilon^2}{4}
g_1(\epsilon),~~\frac{\xi-2\epsilon+\epsilon^2}{6}=
\frac{\epsilon^2}{4}
g_2(\epsilon), \\
g_1(\epsilon)&=& \left(1-\frac{329}{2}\epsilon\right)f,~~
g_2(\epsilon)= 3 \left(1+\frac{247}{36}\epsilon
   -\frac{143}{36}\epsilon^2 \right)f
    +\frac{2}{3} . \nonumber
\end{eqnarray}
Note that $g_i(\epsilon)$'s are of $O(1)$, and thus this shows the
fact that $V_I'$ is order of $\epsilon$ smaller than $\hat H_q$. The
Hamiltonian is then written as,
\begin{eqnarray} \label{H:aq}
\hat{H} &=&
   \frac{\hbar\bar\Omega}{2}\left[\hat{a}_q \hat{a}_q^\dagger
       +\hat{a}_q^\dagger \hat{a}_q
   \right]+\frac{\hbar\Omega_q\epsilon^2}{16}\sum_{i=1}^3g_i(\epsilon)
      \hat{H}_{2i}^0 +\cdots     \\
  &=&\frac{\hbar\bar\Omega}{2}\left[\hat{a}_q \hat{a}_q^\dagger
       +\hat{a}_q^\dagger \hat{a}_q
   \right]+\frac{\hbar\Omega_q\epsilon^2}{16}\left\{
      g_1(\epsilon)
       +3g_2(\epsilon)+15g_3 + \left[g_1(\epsilon)
       +6g_2(\epsilon)+45g_3\right] \sum_{r=0}^2~
        \left(\begin{tabular}{c} 2\\ $r$\end{tabular}\right)
         (\hat{a}_q^\dagger)^{2-r}(\hat{a}_q)^r \right. \nonumber\\
   &+&\left. \left(g_2+15g_3\right)\sum_{r=0}^4~
       \left(\begin{tabular}{c} 4\\ $r$\end{tabular}\right)
       (\hat{a}_q^\dagger)^{4-r}(\hat{a}_q)^r +
       g_3\sum_{r=0}^6~
       \left(\begin{tabular}{c} 6\\ $r$\end{tabular}\right)
        (\hat{a}_q^\dagger)^{6-r}(\hat{a}_q)^r
        \right\} +O(\epsilon^3).  \nonumber
\end{eqnarray}
We now want to write this Hamiltonian as a generalized deformed
oscillator:
\begin{eqnarray} \label{ada:H}
\hat{H}&=&\frac{\hbar \Omega}{2}
\left(\hat{a}_{(1)}\hat{a}_{(1)}^\dagger+\hat{a}_{(1)}^\dagger
\hat{a}_{(1)}\right)+O(\epsilon^3) ,\\
~[\hat{a}_{(1)}, \hat{a}_{(1)}^\dagger]&=& 1+ \epsilon \alpha_1
\hat{a}_{(1)}^\dagger \hat{a}_{(1)}+ \epsilon^2
\alpha_2(\hat{a}_{(1)}^\dagger \hat{a}_{(1)})^2 , \nonumber
\end{eqnarray}
where the algebra defines the deformation function,
$F(y)=1+(1+\epsilon \alpha_1) y+\alpha_2(\epsilon
y)^2$~\cite{bonatsos3}. Since $V_I'$ is $O(\epsilon^2)$, correction
to the annihilation operator would be of order $\epsilon^2$, and
thus, $\hat{a}_{(1)}$ can be written as
\begin{eqnarray} \label{a}
\hat{a}_{(1)}&=& \hat{a}_q+ \epsilon^2\left[\sum_{n=0}^1 u_n
(\hat{a}_{q}^\dagger)^{1-n} (\hat{a}_{q})^{n} +\sum_{n=0}^3 v_n
(\hat{a}_{q}^\dagger)^{3-n} (\hat{a}_{q})^{n}+ \sum_{n=0}^5 w_n
(\hat{a}_{q}^\dagger)^{5-n}
(\hat{a}_{q})^{n} \right] +\cdots . 
\end{eqnarray}
For this operator to satisfy the algebra~(\ref{ada:H}), the
following relations should hold for the coefficients in
Eq.~(\ref{a}):
\begin{eqnarray} \label{uvw}
w_0&=&\frac{2}{15}w_1=\frac{w_2}{24}=\frac{w_3+w_3^*}{20}=
   \frac{g_3}{16}\frac{\Omega_q}{\Omega},~~ w_4=-\frac{w_2}{2},~~
    w_5=-\frac{w_1}{5}, \\
v_0&=&\frac{v_1}{6}= \left(g_2-10g_3\right)\frac{\Omega_q}{16
\Omega}, ~~v_2+v_2^*=\left(g_2+10g_3\right)\frac{3\Omega_q}{8
\Omega},~~v_3=- \frac{v_1}{3}, \nonumber \\
u_1&+&u_1^*=0,~~ u_0=
  \frac{\Omega_q}{16\Omega}(g_1+6g_2+45g_3) , \nonumber
\end{eqnarray}
which give, to $O(\epsilon^2)$,
\begin{eqnarray} \label{Om}
\Omega&=& \left[1+\frac{\epsilon^2}{8} \left(g_1+3g_2+15g_3\right)
\right]\bar\Omega,  \\
\alpha_1 &=&1+\frac{3\epsilon}{4}(g_2+5g_3) ,~~ \alpha_2=
\frac{15}{4}g_3. \nonumber
\end{eqnarray}
To obtain the $1^{\rm st}$ order correction to $O(\epsilon^2)$, we
may set $\Omega_q/\Omega \rightarrow 1$ in Eq.~(\ref{uvw}), since
all the coefficients of $\Omega_q/\Omega$ have factors of order
$O(\epsilon^2)$. The $1^{\rm st}$ order ground state is then defined
by
\begin{eqnarray} \label{0:aa}
\hat{a}_{(1)} |0\rangle_{(1)} =0,
\end{eqnarray}
and its energy is the same as that in Eq.~(\ref{E0:2}), since the
variational approximation leading to (\ref{E0:2}) includes the
contribution from the potential $V_I'$, as in the case of the
Gaussian variational perturbation. The $n^{\rm th}$ excited states
are given by
\begin{eqnarray} \label{n:state}
|n\rangle_{(1)} =\frac{ (\hat{a}_{(1)}^\dagger)^n}{\sqrt{[n]!}}
|0\rangle_{(1)} ,
\end{eqnarray}
where $[n]!=[n][n-1] \cdots [1]$, and $[n]$ is defined by the
recurrence relation $[n+1]=F([n])$, with $[0]=0$. The energy of the
$n^{\rm th}$ eigenstate is given by
\begin{eqnarray} \label{E:n}
E_n= \frac{\hbar \Omega}{2}\left([n]+[n+1]\right) .
\end{eqnarray}
In contrast to the ground state energy that shows a slight
improvement for large $\xi_0$, the energies of the excited states
receive considerable improvements compared to those in the Gaussian
and the post Gaussian approximation, since the expansion parameter
$\epsilon$ is smaller than that of the Gaussian approximation.

The higher order corrections can be obtained similarly as in the
above procedure. To obtain the $2^{\rm nd}$ order correction, for
example, we need to write the Hamiltonian~~(\ref{H:aq}) up to
$O(\epsilon^3)$, and to express it as a function of $\hat{a}_{(1)}$
and $\hat{a}_{(1)}^\dagger$. We then write the Hamiltonian in a
factorized form, $\displaystyle H_{(2)}= \frac{\hbar
\Omega}{2}(\hat{a}_{(2)}\hat{a}_{(2)}^\dagger +\hat{a}_{(2)}^\dagger
\hat{a}_{(2)})+O(\epsilon^4)$, while $\hat{a}_{(2)}$ satisfies a
generalized deformed algebra.

The fact that the first order perturbative result can be expressed
as a generalized deformed oscillator, Eq.~(\ref{ada:H}), also
provides us with a possibility of establishing a new variational
perturbation theory based on the generalized deformed
oscillator~(\ref{ada:H}).

One of the main advantage for using the algebraic approach is that
it enables one to define the thermal state and coherent state
easily. For example, the coherent state can be written as the state
$|\alpha\rangle $,
\begin{eqnarray} \label{coher}
\hat a(t) | \alpha \rangle = |\alpha \rangle \alpha .
\end{eqnarray}
The construction of the state $|\alpha\rangle$ from the number
states can be achieved from the usual method of the
$q-$oscillator~\cite{cho}.

It would be interesting to apply the present method to general
quantum mechanical systems. For example, the realization of
su$_q(2)$ by $q-$boson can be used as a basis of a perturbation
theory for systems with spherical symmetric potential. More
interesting would be to generalize the present method to quantum
field theory by expanding the quantum fields in Fourier modes where
each mode acts as a generalized deformed oscillator.

\vspace{.5cm}
\begin{acknowledgments}
This work was
supported in part by Korea Research Foundation under Project number
KRF-2003-005-C00010 (H.-C.K. and J.H.Y.).
\end{acknowledgments}

\newpage

\begin{appendix}

\section{Calculation of the annihilation operator}

To have an explicit expression for the annihilation operator $\hat
a_q $, we need the $q-$annihilation operator as a function of
$\hat{x}$ and $\hat{p}$,
\begin{eqnarray} \label{a:qdef}
\hat{a}_q= \sum_{l=0} \sum_{n=0}^{2l+1} \epsilon^l u_{n,l}^*
\hat{H}_n^{2l+1-n},
\end{eqnarray}
which satisfies Eqs.~(\ref{H:q}) and (\ref{q:deformed}). In
obtaining this expression, the operator product expansion formula,
\begin{eqnarray} \label{product}
\hat{H}_{m}^{n}\hat{H}_{m'}^{n'}= \sum_{k=0} i^k h_k(m,n;m',n')
\hat{H}_{m+m'-k}^{n+n'-k}
\end{eqnarray}
is useful. The expansion coefficients, $h_k$ are given as
\begin{eqnarray} \label{coefs}
h_0&=&1,~~h_1= m n'-m'n,~~\\
h_2&=&-\left[ m(m-1)+m'(m'-1)\right] n n'-
   m m'\left[n(n-1)+n'(n'-1)\right]- 3 m m'n n', \nonumber \\
h_3(m,n;m',n')&=&2(f_{m}^{n'}-f_{m'}^n)+(f_{m+m'}^n
        -f_m^{n+n'}) -
  3 m'm(m-1)\left(\begin{tabular}{c}
        $n+n'$\\3\end{tabular}\right)
   +3 n'n(n-1)\left(\begin{tabular}{c}
        $m+m'$\\3\end{tabular}\right) \nonumber \\
   && - 3\left[m(m-1)n'(n'-1)-m'(m'-1)n(n-1)\right]
  (m+m'-2)(n+n'-2),  \nonumber
\end{eqnarray}
where $\displaystyle  f_m^n=6\left(\begin{tabular}{c}
$m$\\3\end{tabular}\right)
    \left(\begin{tabular}{c} $n$\\3\end{tabular}\right).$

We assume that the Hamiltonian for the $q$-oscillator is written as
the sum of the kinetic energy and potential term dependent only on
$x$,
\begin{eqnarray} \label{Hq3}
\hat{H}_q &=&\frac{\hat{p}^2}{2}+\frac{\Omega_q^2}{2}x^2+
\frac{c}{4}x^4+ \frac{d}{6}x^6+\cdots .
\end{eqnarray}
If we demand the operator $\hat a_q$ to satisfy the
algebra~(\ref{q:deformed}) and the Hamiltonian~(\ref{H:q}) to be
(\ref{Hq3}), we get $\hat a_q$ in Eq.~(\ref{a:results}) and the
$q-$Hamiltonian
\begin{eqnarray}
\hat H_q&=&\frac{\hbar \Omega_q}{4}\left[\hat{H}_2^0+\hat{H}_0^2+
   \frac{\epsilon}{3}\left(1-\frac{\epsilon}{2}\right)\hat{H}_4^0
   -\frac{g_3\epsilon^2}{4}
     \hat{H}_6^0 +\cdots\right].
\end{eqnarray}
The parameters in~(\ref{H:q}) and (\ref{q-ham}) are determined as
\begin{eqnarray}\label{O:Oq}
\bar{\Omega}&=&\left( 1+\frac{1}{2}\epsilon+\frac{1}{8}\epsilon^2
\right)\Omega_q, ~~c=\frac{4\epsilon \Omega_q^3}{3
\hbar}\left(1-\frac{\epsilon}{2}\right), \\
~~ d&=& - \frac{3 \epsilon^2 g_3 \Omega_q^4}{\hbar^2} , ~~
g_3=-\frac{23}{45}. \nonumber
\end{eqnarray}
The present result can be easily checked using mathematica, in which
the noncommutative multiplication is implemented.

\end{appendix}%


\begin{thebibliography}{10}
\bibitem{bonatsos3} The function $F(y)$ is related to $\Phi(y)$ of
Ref.~\cite{bonatsos} by $\displaystyle F(y)= \Phi(\Phi^{-1}(y)+1)$.
\end{thebibliography}

\vspace{4cm}


\end{document}